\documentstyle[a4wide,12pt,epsf]{article}
\newcommand{\rcite}[1]{{\cite{#1}}}

\newcommand{\tref}[1]{{\ref{#1}}}
\newcommand{\rlabel}[1]{{\label{#1}}}
\newcommand{\rbibitem}[1]{\bibitem{#1}}
\newcommand{\be}{\begin{equation}}
\newcommand{\ee}{\end{equation}}
\newcommand{\ba}{\begin{eqnarray}}
\newcommand{\ea}{\end{eqnarray}}
\begin{document}
\begin{titlepage}
\begin{flushright}
NORDITA 96/45 N,P\\
hep-ph/9607304
\end{flushright}
\vfill
\begin{center}
{\Large\bf CHIRAL PERTURBATION THEORY FOR DA$\Phi$NE PHYSICS}\footnote{Invited
talk presented at the ``Workshop on K Physics,''
Orsay, France, May 30 --- June 4, 1996. To be published in
the proceedings.}\\[2\baselineskip]
Johan Bijnens\\
NORDITA, Blegdamsvej 17,\\
DK 2100 Copenhagen, Denmark\\
\end{center}
\vfill
\begin{abstract}
This talk is an overview of Chiral Perturbation Theory for Kaon physics as
far as not covered by other speakers. It includes a short introduction
to the strengths and weaknesses of Chiral Perturbation Theory
and its applications to semileptonic
Kaon decays and $K\to\pi\pi(\pi)$.
 \end{abstract}
\vfill
\flushleft{July 1996}
\end{titlepage}

\section{Introduction}

The topic of this talk is a review of Chiral Perturbation Theory (CHPT)
for DA$\Phi$NE physics. This title is at the same time too broad and too
narrow for what I will discuss. It is too broad because I will restrict myself
here to the topics related to Kaon decays only and even there I do not discuss
the parts which have been covered by other speakers in this meeting.
It is too narrow because Kaon physics is a very alive subject and is done
at many places besides the experiments at DA$\Phi$NE.

The full explanation of most of what I will present are in ``The Second
DA$\Phi$NE Physics Handbook''\rcite{daphne}. So I have left out from
DA$\Phi$NE physics the following topics
\begin{enumerate}
\item $\gamma\gamma\to\pi\pi$ and $\gamma\gamma\to\pi\pi\pi$.
\item All $\eta$ physics.
\item Hypernuclear physics and Kaon-Nucleon scattering. This is the domain
of the FINUDA detector.
\item Vector meson properties.
\item All topics where CHPT is not relevant.
\end{enumerate}
Even within Kaon physics I have left out all discussion related to
mixings and theoretical estimates of CHPT parameters\rcite{various}. The
$K\to\pi\pi$ aspects have been discussed at length here as well\rcite{epe}
and finally the rare decays were discussed by A.~Pich\rcite{pich}.

In section \tref{semileptonic} I discuss the semileptonic decays and $\pi\pi$
scattering.
$K\to 3\pi$ is treated in section \tref{nonleptonic}.

\section{Chiral Perturbation Theory}

Chiral Perturbation Theory is a systematic way to use the constraints
of chiral symmetry and its spontaneous breakdown. It is a systematic
way to go beyond the PCAC method to reach higher orders in the
expansion in quark masses and momenta. Its basic principles were laid
out by S.~Weinberg in a nicely written paper\rcite{weinberg1}. A systematic
derivation and the use of the external field method then provided the
basis for the revival of these techniques\rcite{GL}.

It is an approach based on:
\begin{itemize}
\item The Chiral Symmetry $SU(3)_L\times SU(3)_R$ of QCD and its spontaneous
breakdown to the vector subgroup $SU(3)_V$.
\item The Goldstone Bosons from this spontaneous breakdown are the only
relevant
degrees of freedom at low energies.
\item Analyticity, causality, cluster expansion and relativity.
\end{itemize}
A proof that these are the only assumptions involved was given by
H.~Leutwyler\rcite{leutwyler}.

So there are only two questions (apart from the technical part of
performing the calculation) when we apply CHPT:
\begin{itemize}
\item Does the expansion in $m_q$ and $p^2$
\footnote{I will use standard CHPT where we count $p^2$ as the same
order as $m_q$. An alternative view can be found
in talk by M.~Knecht and references therein\rcite{knecht}.}, where $p$ is a
generic momentum
or energy, converge ?
\item If yes and the higher orders are important, do we have enough data
to determine all relevant parameters or do we have a sufficiently
reliable way of estimating them ?
\end{itemize}

\section{Semileptonic Decays}
\rlabel{semileptonic}
The lowest order lagrangian is
\be
{\cal L}_2 = \frac{F^2}{4}\mbox{tr}\left(D_\mu U D^\mu U^\dagger
+\chi U^\dagger + U \chi^\dagger\right)\,.
\ee
It contains 2 free parameters, $F$, related to the pion decay constant $F_\pi$,
and $B_0$, which is related to the vacuum expectation value $\langle\bar{q}q
\rangle$, plus the ratios of the quark masses.

The next-to-leading, ${\cal O}(p^4)$, Lagrangian was derived in Ref.\rcite{GL}
and contains an additional 10 parameters labelled $L_i$. All of these are
at present determined from experiment and most can be in principle determined
in Kaon semileptonic decays. In table \tref{table1} the present best values
for these parameters, the source of this value and to which
Kaon decays they contribute are listed. $L_4$ and $L_5$ occur in all decays
but the crosses in table \tref{table1} signify a different combination
than in $K_{l2}$ decays.
\begin{table}
\begin{center}
\begin{tabular}{|c||c|c|c|c|c|c|c|c|c|}  \hline
 & $10^3\cdot L_i(M_\rho)$&Source& & &  &&
\multicolumn{3}{|c|}{$K\to\pi^1\pi^2 e^+\nu$}\\
\cline{8-10}
 &&    &
$K_{l2\gamma}$ &
$K_{l2ll}$    &
$K_{l3}$    &
$K_{l3\gamma}$&
$\pi^+\pi^-$&
$\pi^0\pi^0$&
$\pi^0\pi^-$ \\ \hline
 $L_1$&$0.4\pm0.3$&$K_{e4},\pi\pi$ &      &      &      &      & $\times$
&$\times$ &
\\
 $L_2$&  $1.35\pm0.3$&$K_{e4},\pi\pi$&    &      &      &      & $\times$
&$\times$ &
\\
 $L_3$&$-3.5\pm1.1$&$K_{e4},\pi\pi$ &      &      &      &      &
 $\times$ &$\times$ & $\times$
\\
 $L_4$ &$-0.3\pm0.5$&$1/N_c$&      &      &      &      & $\times$ &$\times$ &
\\
 $L_5$&$1.4\pm0.5$&$F_K/F_\pi$&       &      &      &      & $\times$
 &$\times$ &$\times$
\\
 $L_9$&$6.9\pm0.7$&$r^2_{\pi V}$&      &$\times$&$\times$&$\times$&
 $\times$ &$\times$ &$\times$
\\
 $L_{10}$&$-5.5\pm0.7$&$\pi\to e\nu\gamma$ &$\times$&$\times$&      &$\times$&
     & &
\\
\hline
Anomaly&&&$\times$&$\times$& &$\times$&$\times$&&$\times$
\\
\hline
\end{tabular}
\end{center}
\caption{Occurrence
of the low--energy coupling constants $L_1,\ldots,L_{10}$ and of
the anomaly  in Kaon semileptonic decays.
In $K_{\mu 4}$ decays,
the same constants as in the electron mode (displayed here) occur. In
addition, $L_6$ and $L_8$ enter in the channels
$K^+\rightarrow \pi^+\pi^-\mu^+\nu_\mu$ and
$K^+\rightarrow \pi^0\pi^0\mu^+\nu_\mu.$
\rlabel{table1}  } \end{table}

For more extended discussions and more references see Ref.\rcite{begdaphne}.

\subsection{$K_{l2}$}

The main use of these decays is to determine $F_K$. From the ratio
of the muon to the electron decay we can also test electron-muon
universality. This assumes we can reliably calculate the relevant
electromagnetic radiative corrections. For the absolute value this is
rather unsure but most of the unsure contributions seem to cancel
in the ratio. For a discussion with an optimistic estimate of the error
involved here see Ref.\rcite{finkemeier}. An improvement by a factor of 10
over the present error of about 5\% on this ratio seems feasible from the
statistical point of view.

\subsection{$K_{l2\gamma}$}

This decay has two formfactors depending on the lepton-neutrino mass squared.
The prediction for the axial, $A(W^2)$, and the vector, $V(W^2)$, formfactors
are to order $p^4$\rcite{donoghue,beg}:
\ba
A& =& -\frac{4}{F}\left(L_9 + L_{10}\right) = \frac{-0.030}{m_K}\nonumber\\
V&=&-\frac{1}{8\pi^2 F} = \frac{-0.067}{m_K}\,.
\ea
The form factor $V(W^2)$ is known to order $p^6$\rcite{abbc}.
The correction is
of order 10 to 20\% in the relevant region of phase space and the formfactor
has become $W^2$ dependent to a similar amount.

Present
data are:
\be
m_K\, |A+V| = 0.105\pm 0.008\qquad\mbox{and}\qquad m_K \,|A-V|\leq 0.35\,.
\ee
The $W^2$ dependence
of both form factors has never been measured and data on $A-V$ are rather poor.

\subsection{$K_{l2ll}$}

In these decays there are 3 axial formfactors and one vector one.
The vector one is known to order $p^6$\rcite{abbc}. Most of the rate
and distributions are determined by the three axial form factors. These
are known to order $p^4$\rcite{beg}. Especially in the modes with
$e^+\nu$ in the final state the rate is very much enhanced over the
helicity suppressed Bremsstrahlung part. Extrapolation to full
phase space is rather difficult for the $e^+e^-$ final states due to the
presence of very small invariant masses for the pair. Some
predictions of Ref.\rcite{beg} compared with experiment can be found in table
\tref{table2}. There the effects mentioned are clearly visible.
\begin{table}
\begin{center}
\begin{tabular}{|c||c|c|c|c|}
\hline
$K^+\to$ & $\mu^+ \nu e^+ e^- $&$ e^+\nu e^+e^- $&$
\mu^+\nu\mu^+\mu^-$&$e^+\nu\mu^+\mu^-$\\
\hline
\rule{0cm}{6mm}Experiment & $(1.23\pm0.32)\cdot
10^{-7}$&$(2.8^{+2.8}_{-1.4})\cdot 10^{-8}$&
$\leq 4.1\cdot10^{-7}$ & $-$ \\
Tree, cuts & $4.98\cdot 10^{-8} $&$2.1\cdot10^{-12}$&$3.8\cdot10^{-9}$&
$3.1\cdot10^{-12}$\\
1-loop, cuts & $8.5\cdot10^{-8}$&$3.4\cdot10^{-8}$&$1.35\cdot10^{-8}$
&$1.1\cdot10^{-8}$\\
1-loop, full & $2.49\cdot10^{-5}$&$1.8\cdot10^{-7}$&$1.35\cdot10^{-8}$&$
1.1\cdot10^{-8}$\\
\hline
\end{tabular}
\end{center}
\caption{\rlabel{table2}Branching ratios from experiment and theory. The cuts
are $m_{e^+e^-}\geq 140~MeV$. No cuts for the decays with $\mu^+\mu^-$.
The last row is with integrating over all of phasespace.}
\end{table}

\subsection{$K_{l3}$}

This was calculated to order $p^4$ by Gasser and Leutwyler\rcite{GL2}.
There are two form factors here which for $K^{+,0}\to\pi^{0,-} l^+ \nu_l$
are parametrized as
\be
\langle\pi(p')|V_\mu^{4-i5}|K(p)\rangle =
\frac{1}{\sqrt{2}}\left[
(p' + p)_\mu f_+(t) + (p-p')_\mu f_-(t)\right]\,,
\ee
with $t = (p-p')^2$. In the analysis of experimental data usually one
uses instead of $f_-$ the scalar formfactor $f_0$.
\be
f_0(t) = f_+(t) + \frac{t}{m_K^2-m_\pi^2} f_-(t)\,.
\ee
These are both parametrized in a linear fashion.

For $f_+$ the $p^4$ expression
fits the linear parametrization very well over the relevant range. Using
$f_+(t) = f_+(0) [ 1 + \lambda_+ t / m_\pi^2]$ theory and experiment agree
very well. Data give $\lambda_+ = 0.029\pm0.002$ while $p^4$ predicts from
the value of $L_9$ in table \tref{table1}, $\lambda_+ = 0.031$. Even the
isospin breaking predicted from CHPT is observed in the measured values
of $f_+(0)$ for both decays.

For $f_0$ the linear parametrization $f_0(t) = f_+(0) [1+\lambda_0 t/m_\pi^2]$
fits satisfactorily the $p^4$ expression. The prediction depends
only on $F_K/F_\pi$ and is $\lambda_0 = 0.017\pm0.004$. The experimental
situation needs clarification. Trying to average incompatible measurements
leads to $0.025\pm0.006$ for $K^0_{\mu3}$ and $0.004\pm0.007$ in $K^+_{\mu3}$.
In addition to the interest for the strong interaction effects these decays
are our major source of knowledge of $|V_{us}|$.

\subsection{$K_{l3\gamma}$}

This process was calculated to order $p^4$ in Ref.\rcite{beg}. There
are 10 formfactors possible here which are all nonzero at order $p^4$.
The calculations shows a rather complicated interplay between all the
various contributions. In the rates the final effect is rather small.
Various distributions might show more sensitivity to the higher order
contributions. In other processes like $K_{l4}$ there are large corrections
so the prediction of small corrections to the rate is definitely nontrivial.
As an example with cuts $E_\gamma\geq 30~MeV$ and $\theta_{e\gamma}\geq 20^o$
we obtained for the branching ratio for $K^+_{e3\gamma}$
\be
\mbox{tree } 2.8\cdot10^{-4}
\quad\stackrel{\mbox{+ }L_i}{\longrightarrow}\quad
3.2\cdot10^{-4}
\quad\stackrel{\mbox{+ loops}}{\longrightarrow}\quad
3.0\cdot10^{-4}\,.
\ee
Afterwards there was a new result from NA31\rcite{NA31} for $K_{Le3\gamma}$
with the same cuts
\be
BR = (3.61\pm0.14\pm0.21)\cdot10^{-3}
\ee
to be compared with\rcite{beg}
\be
\mbox{tree } 3.6\cdot10^{-3}
\quad\stackrel{\mbox{+ }L_i}{\longrightarrow}\quad
4.0\cdot10^{-3}
\quad\stackrel{\mbox{+ loops}}{\longrightarrow}\quad
3.8\cdot10^{-8}\,.
\ee
These are in excellent agreement. All possible decay modes should in fact be
observable in the near future so the other predictions will also be tested.

\subsection{$K_{l4}$}

In these decays there are 4 formfactors possible. They all have a quite
different behaviour under chiral perturbation theory. The four form factors
are 3 axial ones, $F$, $G$ and $R$, and one vector one $H$. They
contribute to the various decays as shown in table \tref{table3}. The other
form factors also contribute but are very small.
\begin{table}
\begin{center}
\begin{tabular}{|l|c|c|}
\hline
Decay & $ l = e$ & $ l = \mu$\\
\hline
$K^+\to\pi^+\pi^-l^+\nu$ & $F,G,H$ & $ F,G,R,H$ \\
$K^+\to\pi^0\pi^0l^+\nu$ & $F$ & $F,R$ \\
$K^0\to\pi^0\pi^-l^+\nu$ & $G,H$ & $G,H,R$ \\
\hline
\end{tabular}
\caption{\rlabel{table3}The contributions of the various formfactors to the
different $K_{l4}$ decays.}
\end{center}
\end{table}
Good measurements at present exist for the formfactors mainly of $K_{e4}$
only. $F,G$ were calculated to $p^4$ in Refs.\rcite{kl4} and $R$ in
Ref.\rcite{BCG}. The formfactor $H$ is known to order $p^6$\rcite{abbc}.
In Ref.\rcite{BCG} higher order effects were also estimated using dispersion
relations. Conclusions of these papers were:
\begin{itemize}
\item $F$ and $G$ get large corrections from their tree level value of
$m_K/(\sqrt{2} F_\pi)$ and allow for an accurate measurement of $L_1$, $L_2$
and $L_3$. These then allow for a clean prediction of the total
rates\rcite{BCG}
reproduced in table \tref{table4}. The errors in the predictions are in fact
dominated by the most accurate measurement now available\rcite{rosselet}.
\begin{table}[htb]
\begin{center}
\begin{tabular}{|c|cccccc|}
\hline
$\pi\pi$ charge & $+-$ & $00$ & $+-$ & $00$ & $0-$ & $0-$ \\
Leptons & $e^+\nu$ & $e^+\nu$ & $\mu^+\nu$ & $\mu^+\nu$ &$e^+\nu$ &
$\mu^+\nu$\\
\hline
Tree & 1297 & 683 & 155 &  102 & 561 & 55\\
$p^4$ & 2447 & 1301 & 288 & 189 & 953 & 94 \\
Full & input & 1625(90) & 333(15) & 225(11) & 917(170) & 88(22)\\
\hline
Exp. & 3160(140) & 1700(320) & 1130(730) & $-$ & 998(80) & $-$\\
\hline
\end{tabular}
\end{center}
\caption{\rlabel{table4}Predictions from Ref.\protect{\rcite{BCG}}
for the various $K_{l4}$ decay widths. The
last two columns are normalized to $K_L$ decays. Full includes the
unitarization
estimates. Exp. are the experimental values. Errors are in brackets and all
values are in $s^{-1}$.}
\end{table}
There is excellent agreement with all available experimental results and
several predictions remain to be tested.
\item $R$ is a relatively small effect even in $K_{\mu4}$ decays. With
accurate measurements of $F$ and $G$ in $K_{e4}$ and distributions in
$K_{\mu4}$
a good determination might still be possible. This would allow a test of the
$1/N_c$ assumption used
in setting $L_4$ essentially to zero. For other possible
relevance of measurements of $R$ see Ref.\rcite{knecht}.
\item $H$ has relatively small higher order corrections if $F_\pi$ is used in
its $p^4$ expression. The slope also is very small\rcite{abbc}.
\end{itemize}
This was all for the real part of the formfactors essentially. The imaginary
part allows us to extract more information, see the next subsection.

\subsection{$\pi\pi$ scattering}

$K_{l4}$ decays also allow an accurate measurement of some $\pi\pi$
scattering angles. These are now known to order $p^6$ or two-loops
in chiral perturbation theory\rcite{pipius} and in generalized
CHPT\rcite{pipiknecht}. They can be easily obtained using the Pais-Treiman
asymmetry methods\rcite{cks}. A comparison of the calculation of
Ref.\rcite{pipius} with the present data
 is shown in Fig. \tref{fig1}. $\pi\pi$ scattering will
allow for a clean test of generalized versus standard CHPT. The
generalized CHPT result allows for a somewhat larger range of scattering
angles. That allows for results between our 2-loop calculation and a curve
roughly following the top of the last three error bars in fig. \tref{fig1}.
\begin{figure}
\begin{center}
\leavevmode\epsfxsize=14cm\epsfbox{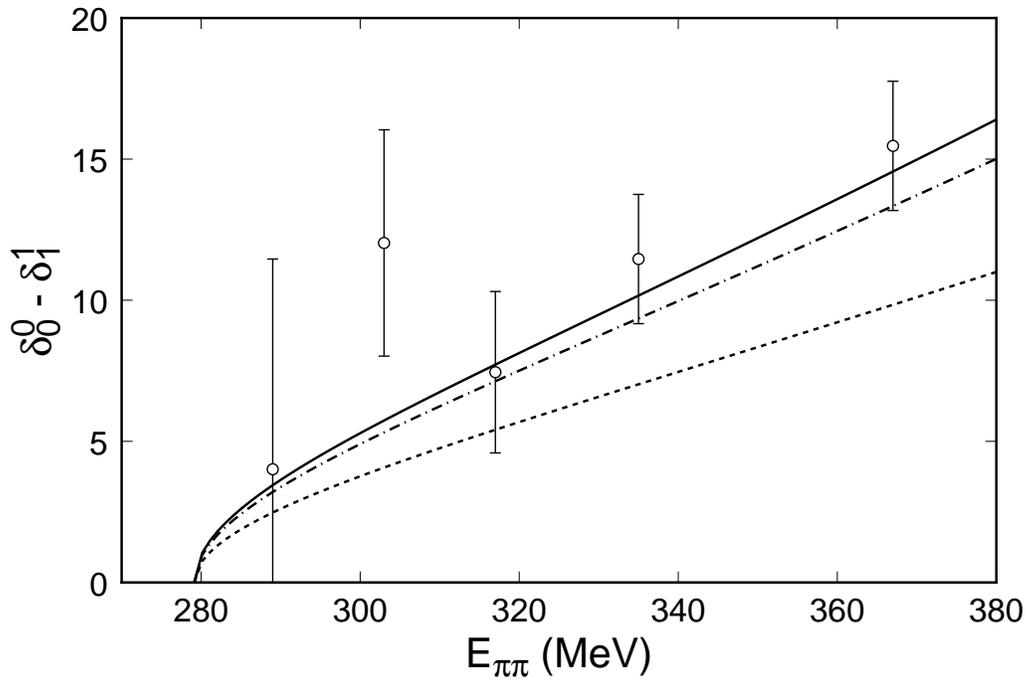}
\end{center}
\caption{\rlabel{fig1}The combination of $\pi\pi$ phase shifts that can be
measured in $K_{l4}$ decays at two loops in CHPT\protect{\rcite{pipius}}.
Also shown are the present data\protect{\rcite{rosselet}}. The curves are:
tree level (dashed), $p^4$ (dot-dashed) and $p^6$ (full).}
\end{figure}

\section{Nonleptonic Decays}
\rlabel{nonleptonic}

A more extended version of the present discussion is present in Ref.
\rcite{MP}.
The main CHPT calculation in this respect is Ref.\rcite{k3pi}.
The number of free parameters in the weak chiral lagrangian is rather
large but still relevant predictions can be made. To lowest order, $p^2$,
there are two parameters, $c_2$ and $c_3$, that are essentially the strength
of the octet and 27 transitions.
At $p^4$ we now restrict to those in $K\to\pi\pi(\pi)$ decays.
For the leading octet parameters
to order $p^4$ there are 3 more. One of these can not be disentangled
from $c_2$ within these decays. For the 27 there are $4$ more one of which
appears always in the same combination with $c_3$ in these decays and is
hence also not relevant. The total number of parameters is thus 7.

The number of observables in the CP-conserving part is two rates (notice I
assume isospin throughout) for $K\to2\pi$ and 14 in the $K\to3\pi$ rates
and Dalitz plots. The latter are using a quadratic representation of
the Dalitz plot and a linear one for the phases:
\begin{tabbing}
\qquad\=Constant part :\quad\= 1(Octet) + 1(27) + 1(phase) \\
\>Linear :  \>1(Octet) + 2(27) + 3(phases)\\
\>Quadratic: \> 2(Octet) + 3(27)
\end{tabbing}
The phases in principle should be calculable in CHPT since they are at very low
$\pi\pi$ center of mass energies. They have not been measured at present.
In table 5 I give the list of parameters, see Ref.\rcite{MP} for their
definition, the present experimental values, the CHPT tree level predictions
and the $p^4$ tree level predictions.
The $K\to2\pi$ rates have always been used as input.
\begin{table}
\begin{center}
\begin{tabular}{|c|ccc|}
\hline
Variable & Tree & $p^4$ & Exp. \\
\hline
$\alpha_1$ & 74 & input   &$91.71\pm0.32$\\
$\beta_1$ &$-16.5$&input &$-25.68\pm0.27$\\
$\zeta_1$&&$-0.47\pm0.18$&$-0.47\pm0.15$\\
$\xi_1$ &&$-1.58\pm0.19$&$-1.51\pm0.30$\\
$\alpha_3$& $-4.1$&input&$-7.36\pm0.47$\\
$\beta_3$&$-1.0$&input&$-2.43\pm0.41$\\
$\gamma_3$&1.8&input&$2.26\pm0.23$\\
$\xi_3$&&$0.092\pm0.030$&$-0.12\pm0.17$\\
$\xi^\prime_3$&&$-0.033\pm0.077$&$-0.21\pm0.51$\\
$\zeta_3$&&$-0.011\pm0.006$&$-0.21\pm0.08$\\
\hline
\end{tabular}
\end{center}
\caption{\rlabel{table5}The parameters of the Dalitz plot in $K\to3\pi$ decays.
The first 4 are octet and the last 6 are 27. The quadratic ones are not
present at tree level. Experimental values are taken from Ref.
\protect{\rcite{k3pi}}.}
\end{table}

Notice that there is in fact a direct
relation between several of the parameters predicted in CHPT. Counting the
number of observables and parameters there should be 5 relations. These are:
\ba
\mbox{Octet :}&&\alpha_1 \to \zeta_1\nonumber\\
&&\beta_1 \to \xi_1\nonumber\\
\mbox{27 :}&&\alpha_3\to\zeta_3\nonumber\\
&&\beta_3\to\xi_3\nonumber\\
&&\gamma_3\to\xi^\prime_3\nonumber
\ea
These are clean predictions of CHPT and should be more stringently tested.
The measurements, especially in the smaller 27 sector, should be relatively
easy to improve using the new facilities. The agreement at present is very
good within the errors.

The CP-violating asymmetries in the Dalitz plot are expected to be of
order $10^{-6}$. The main reason for this is that in order for the CP phase
to be observable it has to interfere with the final state phases. These
are small since the pions in $K\to3\pi$ are at very low energies. In addition
to order $p^4$ the interference happens only with the suppressed 27 amplitudes.
So at order $p^4$ one expects asymmetries of order $10^{-6}$.
The final number might be significantly enhanced by $p^6$ effects
where interference with the dominant octet amplitudes becomes possible.
A CHPT inspired estimate of this effect is in Ref.\rcite{detal}.

\section{Conclusions}

Chiral Perturbation Theory for Kaons is in very good shape as can be seen
from this talk and various others in this meeting.

In the
semileptonic sector all parameters to order $p^4$ are determined and
various good tests have already been obtained and we look forward to
more tests in the near future. On the theoretical side the push beyond
$p^4$ has slowly started, e.g. in $\pi\pi$ scattering, and more data are
very welcome.

In the nonleptonic sector it has so far been most powerful in
rare decays\rcite{pich}. Restricting to $K\to2\pi$ and
$K\to3\pi$ the present experimental tests of the $p^4$ relations
are only relevant in the octet part. Chiral symmetry does however provide
a simple explanation for the various sizes of the Dalitz plot parameters,
with the exception of $\Delta I=1/2$ rule. We look forward to more
stringent tests here in the future.

\section*{Acknowledgements}

I would like to thank the organizers and especially L.~Fayard for a well
organized and very nice meeting. I would also like to thank
Ll. Ametller, A. Bramon, G. Colangelo, F. Cornet, G. Ecker, J. Gasser and
M. Sainio for a series of pleasant collaborations.

\end{document}